\documentclass[english,keywords,showpacs,amsmath,amssymb,showpacs]{iopart}                                                       
\usepackage{graphicx}
\usepackage{color}
\usepackage{bm}
\usepackage{longtable}
\begin{document} 
\title{Using the no-signaling condition for constraining the nonidealness of a Stern-Gerlach setup}
\author{Dipankar Home\footnote{dhome@bosemain.boseinst.ac.in}$^1$
and Alok Kumar Pan\footnote{apan@bosemain.boseinst.ac.in}$^1$}

\address{$^1$ CAPSS, Department of Physics, Bose Institute, Salt Lake, Calcutta
700091, India}
\begin{abstract}
On the basis of a variant of the EPR-Bohm example, we show that the no-signaling condition can be employed as a useful tool for deriving a constraint on a suitably defined measure of the `nonidealness' of a Stern-Gerlach(SG) setup.  In this demonstration, a key ingredient is provided by the characteristics of the exact solution of the time-dependent Schroedinger equation as applied to a most general SG setup.
\end{abstract}
\maketitle
\section{Introduction}
Spurred on by Bell's seminal work[1] related to the EPR-Bohm example[2,3], the study of quantum mechanical correlations  between the results of measurements on  the spatially separated particles in the entangled states has become a vibrant research enterprise. Among its various ramifications,  a particularly curious feature is that, in spite of an underlying `nonlocality' embodied in the EPR-Bohm type correlations, the rules of quantum mechanics turn out to ensure that such `nonlocality' cannot be used for sending information in a controlled way that may lead to causality paradoxes. The way this no-signaling condition(`signal locality') is satisfied by the quantum mechanical formalism for the entangled states has already been the subject of a number of analyses in different forms[4-10]; nevertheless, it is instructive to probe with respect to new types of examples the way the validity of this condition gets ensured, thereby leading to interesting constraints on the operations of certain quantum devices.

It is in the above mentioned context that we probe in this paper a variant of the EPR-Bohm example that has a special interest because it involves the use of a \emph{nonideal} quantum measurement(viz. by using the most general nonideal Stern-Gerlach(SG) setup) in which the properties  of explicit solutions of the relevant time-dependent Schroedinger equation play a critical role. Then, in the example considered here, we find that even though the Schroedinger dynamics is intrinsically nonrelativistic,  the relevant specifics of the Schroedinger dynamics turn out to be compatible with the no-signaling condition, crucially through  a mathematically  valid inequality that acts as a \emph{constraint} limiting an appropriately defined measure of the `nonidealness' of a SG setup. Before demonstrating this result in Section 3 on the basis of an appropriate physical reasoning, we first formulate in the next section the required variant of the EPR-Bohm example, alongside delineating some key features of a nonideal SG setup that will be used in  our argument.
\section{The EPR-Bohm example with a nonideal SG setup}
Let us begin with a source emitting EPR-Bohm entangled pairs in spin singlets. In particular, we consider the pairs propagating along opposite directions. The initial total wave function is given by
\begin{equation}
|\Psi\rangle_{i}=\frac{1}{\sqrt{2}}|\psi_{0}\rangle_{1}|\psi_{0}\rangle_{2}\left(|\uparrow\rangle_{1}|\downarrow\rangle_{2} -|\downarrow\rangle_{1}|\uparrow\rangle_{2}\right)
\end{equation}
where the spatial parts $|\psi_{0}\rangle_{1}$ and $|\psi_{0}\rangle_{2}$(represented by Gaussian wave packets) correspond to particles 1 and 2 respectively , and the spin part corresponds to the singlet state. 

Next, a SG setup be placed along one of the two wings of the EPR-Bohm pairs, say,  for particles $2$ moving along the $+y$-axis(Figure 1). After passing through the inhomogeneous magnetic field in the SG setup oriented along, say, the $+z-axis$, these particles  belong to the spatially separated wave packets represented by $|\psi_{+}({\bf x},t)_{2}|^{2}$ and $|\psi_{-}({\bf x},t)_{2}|^{2}$. These two wave packets freely propagate along the y-z plane (with equal and opposite momenta of their peaks), corresponding to spin up $(\left|\uparrow\right\rangle)$ and spin down $(\left|\downarrow\right\rangle)$ states respectively.

Consequently, the total wave function of the entangled pairs after the particles 2 have gone through the SG magnetic field is given by
\begin{equation}
|\Psi\rangle_{SG}=\frac{1}{\sqrt{2}}|\psi_{0}\rangle_{1}\left[\psi_{-}({\bf x},t)_{2}\left|\uparrow\right\rangle_{1}\left |\downarrow\right\rangle_{2}-\psi_{+}({\bf x},t)_{2}\left|\downarrow\right\rangle_{1}\left|\uparrow\right\rangle_{2}\right]
\end{equation}
where $\psi_{+}({\bf x},t)_{2}$ and $\psi_{-}({\bf x},t)_{2}$ are solutions of the time-dependent Schrödinger equation for the SG setup, containing the interaction term $H_{int}=\mu\widehat{\sigma}.B$. The explicit forms of the spatial wave functions $\psi_{+}({\bf x},t)_{2}$, $\psi_{-}({\bf x},t)_{2}$, and the details of the  relevant mathematical treatment are given in the Appendix, where the corresponding initial wave function of particles 2 is taken to be of the following form
\begin{equation}
\psi_{0}\left({\bf x},0\right)_{2}=\frac{1}{{(2\pi\sigma_{0}^{2})}^{{3}/{4}}}
\exp\left(-
\frac{{\bf x}^{2}}{4\sigma_{0}^{2}}+ i k y\right)
\label{initpack}
\end{equation}
Here the wave packet $|\psi_{0}({\bf x},0)_{2}|^{2}$ is peaked at the entry point($x=y=z=0$) of the SG magnetic field region, and $\sigma_{0}$ is the initial width of the wave packet. 
\vskip -3cm
\begin{figure}[h]
{\rotatebox{0}{\resizebox{14.0cm}{11.0cm}{\includegraphics{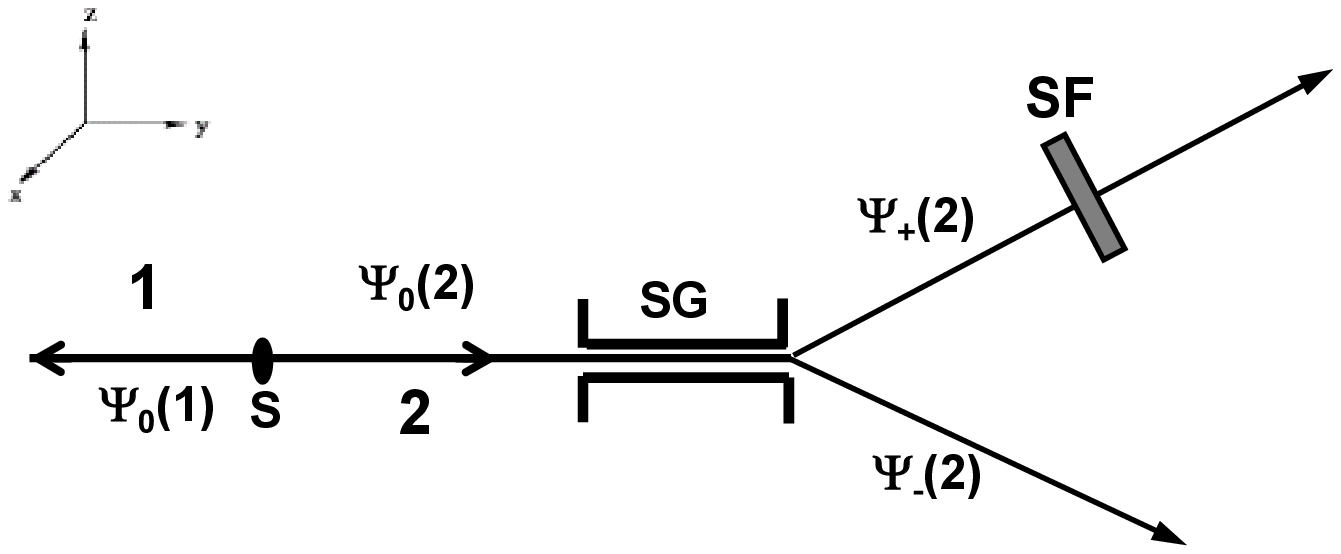}}}}
\end{figure}
\vskip -4cm
{\footnotesize Figure 1. A schmetic setup  of the EPR-Bohm type example using Stern-Gerlach device in one of the two wings of the entangled pairs of particles(see text for details).}\\

Now, we focus on the subensemble of particles 2 emerging from the SG setup that are confined to the upper y-z plane ($y=0$ to $+\infty$ and $z=0$ to $+\infty$).  These particles are selected out and passed through a spin-flipper(SF) which flips the spin state $\left |\uparrow\right\rangle_{2}$ to $\left |\downarrow\right\rangle_{2}$.  The combined state of particles 1 and 2 after this operation is given by
\begin{equation}
|\psi\rangle_{SG+SF}= \frac {1}{2}|\psi_{0}\rangle_{1}\left[|\psi_{-}({\bf x},t)\rangle_{2}|\uparrow\rangle_{1}|\downarrow\rangle_{2} - |\psi_{+}({\bf x},t)\rangle_{2}|\downarrow\rangle_{1}|\downarrow\rangle_{2}\right]
\end{equation}

Examining the validity of the no-signaling condition in this case boils down to probing under what condition the expectation value of an arbitrary spin observable pertaining to the particles 1 remains unaffected by the above mentioned spin-flipping operation on the subensemble of particles 2. For this, we have to first pinpoint the relevant features of a \emph{nonideal} SG setup, clarifying precisely the criteria of `ideal' and `nonidealness' of a SG setup.

The usual description of an \emph{ideal} measurement of spin (in this particular case, of the variable $\widehat{\sigma}_{z}$) using the SG setup assumes the following conditions to be satisfied:

${\bf A.}$ The  wave functions $\psi_{+}({\bf x},t)_{2}$ and $\psi_{-}({\bf x},t)_{2}$ are mutually orthogonal. This means that the configuration space distinguishability between the wave functions $\psi_{+}({\bf x},t)_{2}$ and $\psi_{-}({\bf x},t)_{2}$ defined in terms of the modulus of their inner product is vanishingly small; i.e., 
\begin{equation}
I=\left|\int_{-\infty}^{+\infty}\psi^{*}_{+}({\bf x},t)_{2} \psi_{-}({\bf x},t)_{2} d^{3}{\bf x}\right|\approx 0
\end{equation}

${\bf B.}$ The probability of finding  particles with the spin $|\uparrow\rangle_{2}(|\downarrow\rangle_{2})$ in the lower(upper) y-z plane is vanishingly small. Satisfying this condition means that the wave packets $|\psi_{+}({\bf x},t)_{2}|^{2}$ and $|\psi_{-}({\bf x},t)_{2}|^{2}$ emerging from the SG setup get well separated in position space(i.e., they eventually become macroscopically distinct) so that the following conditions hold good: 
\begin{equation}
\fl
|\alpha|^{2}=\int_{x=-\infty}^{+\infty}\int_{y=0}^{+\infty}\int^{0}_{z=-\infty}|\psi_{+}({\bf x},t)_{2}|^{2}d^{3}{\bf x}=\int_{x=-\infty}^{+\infty}\int_{y=0}^{+\infty}\int_{z=0}^{+\infty}|\psi_{-}({\bf x},t)_{2})|^{2}d^{3}{\bf x}\approx 0
\end{equation}
and 
\begin{equation}
\fl
|\beta|^{2}=\int_{x=-\infty}^{+\infty}\int_{y=0}^{+\infty}\int^{0}_{z=-\infty}|\psi_{-}({\bf x},t)_{2}|^{2}d^{3}{\bf x}=\int_{x=-\infty}^{+\infty}\int_{y=0}^{+\infty}\int_{z=0}^{+\infty}|\psi_{+}({\bf x},t)_{2}|^{2}d^{3}{\bf x}\approx 1
\end{equation}

At this stage, it is important to note that, depending upon the choices of the relevant parameters(viz. the magnetic field gradient, the interaction time, and the initial width of the wave packet), the validity of the condition ${\bf A}$ does \emph{not} automatically ensure the validity of the condition ${\bf B}$  - the latter is, in fact, operationally the key condition for the `idealness' of the SG setup when it is used for spin measurement\cite{home08}. Hence for defining, in general, a nonideal SG setup, the question of violation of the condition ${\bf B}$ plays a crucial role. 

Note that the quantity $I$ remains unchanged with time after the relevant wave packets $|\psi_{+}({\bf x},t)_{2}|^{2}$ and $|\psi_{-}({\bf x},t)_{2}|^{2}$ emerge from the SG magnetic field region. This is because these two wave packets evolve under the \emph{same} unitary evolution; i.e., both of them  move freely with equal and opposite momenta of their peaks . In contrast, the quantities  $|\alpha|^{2}$ and $|\beta|^{2}$ are time dependent and, interestingly,  both of them \emph{saturate} to  \emph{time independent constant values} at a certain time after emerging from the SG setup(see Appendix for the details of how this `saturation' occurs as a consequence of a rigorous solution of the relevant time-dependent Schroedinger equation). 

Thus, in order to appropriately characterize the most general \emph{nonidealness} of a SG setup, it is necessary to use a measure of the `nonidealness' that can encapsulate the features associated with the parameters $|\alpha|^{2}$,$ |\beta|^{2}$ given by Eqs.(6) and (7). A convenient choice for this purpose is a quantity defined in the following way
\begin{equation}
M(t)=\int_{-\infty}^{+\infty}\sqrt{|\psi^{*}_{+}({\bf x},t)_{2}|^{2}\:| \psi_{-}({\bf x},t)_{2}|^{2}} \:d^{3}{\bf x}
\end{equation}
which denotes the position space overlap between the oppositely moving wave packets corresponding to $|\psi_{+}({\bf x},t)_{2}|^{2}$ and $| \psi_{-}({\bf x},t)_{2}|^{2}$.

Note that the parameter $M(t)$  is, in general,  time dependent, but \emph{saturates} to a time independent value $M_{s}$ after a certain time $t=t_{s}$ depending upon the values of the relevant parameters in the SG setup. The  lower and upper bounds of $M_{s}$ are $0$ and $1$ respectively.  

Hence, for an ideal SG setup we have $I\approx 0$ and $M_{s}\approx 0$. But, by choosing the values of the relevant parameters one can ensure nonzero appreciable values of both the quantities $I$ and $M_{s}$. Therefore, the most \emph{general} type of \emph{nonideal} SG setup is characterized by the conditions $I\neq0$ and $M_{s}\neq0$. Then, in the context of the EPR-Bohm setup, the following question immediately suggests itself: Is there a relationship between  $I$ and $M_{s}$, or, a bound to the value of $M_{s}$ in  a most general nonideal SG setup that can be related to the no-signaling condition? 

That such a constraint can indeed be obtained is demonstrated in this paper by showing that $M_{s}$ has to be always greater than or equal to $I$; otherwise the no-signaling condition(signal locality) would be violated. 

\section{Constraint on the SG nonidealness from the no-signaling condition}
In the  variant of the EPR-Bohm example we are considering using the most general \emph{nonideal} SG setup, a subensemble of particles 2 emerging from the SG magnetic field that are confined to the upper y-z plane(i.e.,  $y\rightarrow 0$ to $y\rightarrow +\infty$ and $z=0$ to $z\rightarrow+\infty$) are selected out and passed through a spin-flipper(SF). Given this scenario, our analysis proceeds as follows.

In the first stage of the argument, we consider what would happen if the relevant parameters could be adjusted such that inner product($I$) has a \emph{nonzero} value, while the position space overlap $M_{s}$ is \emph{vanishingly small}, i.e., $I\neq0$ but $M_{s}\approx0$. Such a condition would operationally mean that a negligibly small number of  particles with spin $|\downarrow\rangle_{2}$($|\uparrow\rangle_{2}$) will be present in the upper(lower) y-z plane. Now, suppose in such a situation, the particles 2 in the upper y-z plane are selected out and passed through the SF, having their spin state  $|\uparrow\rangle$ flipped to the spin state $|\downarrow\rangle$.

Then, the expectation value of an arbitrary spin observable $A$ pertaining to the particles 1 in the other wing of the EPR-Bohm pairs can be written as follows by using Eq.(4)
\begin{eqnarray}
\fl
\nonumber
\left\langle\psi|A|\psi\right\rangle_{SG+SF}&=& \frac{1}{2}\left[\langle\uparrow|A|\uparrow\rangle_{1}|\psi_{-}({\bf x},t)_{2}|^{2}+\ \langle\downarrow|A|\downarrow\rangle_{1}|\psi_{+}({\bf x},t)_{2}|^2\right]\\
&&+\left[_{2}\langle\psi_{-}({\bf x},t)|\psi_{+}({\bf x},t)\rangle_{2}\  \langle\uparrow|A|\downarrow\rangle_{1} \right]
\end{eqnarray}

On the other hand, without subjecting the particles 2 in the upper y-z plane to the spin-flipping operation, the expectation value of the above observable, as calculated by using Eq.(2), is given by  
\begin{eqnarray}
\fl
\left\langle\psi|A|\psi\right\rangle_{SG}= \frac{1}{2}\left[\langle\uparrow|A|\uparrow\rangle_{1}|\psi_{-}({\bf x},t)_{2}|^{2}+\ \langle\downarrow|A|\downarrow\rangle_{1}|\psi_{+}({\bf x},t)_{2}|^2\right]
\end{eqnarray}

Hence, it is evident from Eqs.(9) and (10) that in the supposed situation where the quantity $M_{s}$ could be considered negligbly small($\approx 0$), along with a finite nonzero value of $I$,  there would be a violation of the no-signaling condition - a violation that may be quantified by a parameter defined in the following way  
\begin{eqnarray}
\Delta=\left[\left\langle\psi|A|\psi\right\rangle_{SG+SF}- \left\langle\psi|A|\psi\right\rangle_{SG}\right]= I [\langle\uparrow|A|\downarrow\rangle_{1}]
\end{eqnarray}
It then follows that the maximum value of $\Delta$ would be given by $\Delta_{max}=I$. 

Here note that, in order to be compatible with the no-signaling condition, the realizability of a nonideal SG setup where  $I\neq0$ and $M_{s}\approx0$ must be \emph{ruled out}. Next, considering the most general nonideal situation where both the quantities $I$ and $M_{s}$ are appreciably  nonzero (i.e., $I\neq0$  and $M_{s}\neq0$), let us examine whether a bound to the value of $M_{s}$ can be obtained from the no-signaling condition. For this, we adopt the following strategy for formulating the relevant argument.

To start with, we note that the above estimation of the value of $\Delta_{max}$ by taking $M_{s}\approx0$ is obviously erroneous if the quantity $M_s$ has a non-negligible value. Then a plausible measure of the error involved in such an estimation is the position space overlap parameter itself. Consequently,  for the no-signaling condition to be satisfied, this error \emph{cannot} be smaller than the value of $\Delta_{max}$. This lends itself to the implication that there has to be a \emph{lower bound} to the value of $M_{s}$, i.e., for all possible choices of the relevant parameters, any given nonideal SG setup with finite nonzero values of both the quantities $M_{s}$ and $I$ must satisfy the following inequality 
\begin{equation}
M_{s}\geq I
\end{equation}                                  
In other words, the validity of the no-signalling condition in this situation gets related to the following mathematical inequality 
\begin{equation}
\int_{-\infty}^{+\infty}|\psi_{+}({\bf x},t)_{2}||\psi_{-}({\bf x},t)_{2}|d^{3}{\bf x}\geq \left|\int_{-\infty}^{+\infty}\psi^{*}_{+}({\bf x},t)_{2} \psi_{-}({\bf x},t)_{2} d^{3}{\bf x}\right|
\end{equation}
We stress that this is the most \emph{general constraint} on the `nonidealness' of the SG setup that can be obtained from the no-signaling condition. While the mathematical validity of the above inequality can be seen from the properties of complex functions, it is interesting that the physical condition of compatibility with the no-signaling condition in the EPR-Bohm example warrants the inevitability of such a constraint. 

Note that, as a particular case, it follows that the no-signaling condition rules out the realizability of a SG setup such that $I\neq 0$ and $M_{s}\approx0$. There is, of course, another particular case of `nonidealness', namely, corresponding to $I\approx0$ and $M_{s}\approx0$ which, obviously, does not lead to any inconsistency with the no-signaling condition, as is evident from  Eq.(13).
\section{Concluding Remarks}
The power of the no-signalling condition in giving rise to specific bounds on various types of quantum operations, such as the limits on the fidelity of quantum cloning machines, and on quantum state discriminations have been demonstrated in a number of ways\cite{gisin}. Our present work complements these studies from a somewhat different perspective, namely, by linking the no-signalling condition with a constraint relation governing an archetypal example of quantum measurement of spin provided by the SG setup. This suggests the possibility of more uses of the no-signalling condition for probing the bounds inherent in the quantum mechanical modelling of other specific measurement processes - a line of investigation which, in conjunction with the studies made to obtain from the no-signalling condition limits on the possible extensions of quantum mechanics\cite{simon}, may lead to some interesting restrictions on generalisations of the quantum theory of measurement. This is currently being studied.

\appendix
\section{}
For the sake of completeness, here we give a concise presentation of the ingredients of the quantum mechanical treatment of the SG setup as relevant to our present paper, while the analyses of the SG setup have been discussed in various contexts\cite{bohmbook, home08,scully}. 
A beam of \emph{x-polarized} spin-1/2 neutral particles, say, neutrons, passing through the SG magnetic field is represented by the total wave function 
$\Psi\left(\textbf{x},t=0\right)=\psi_{0}\left(\textbf{x}\right)\chi(t=0)$.  The spatial part $\psi_{0}({\bf x})$ 
corresponds to a Gaussian wave packet which is initially peaked at the entry point(${\bf x}=0$) of the SG magnet at $t=0$, given by
\begin{equation}
\psi_{0}\left({\bf x}\right)=\frac{1}{{(2\pi\sigma_{0}^{2})}^{{3}/{4}}}
\exp\left(-
\frac{{\bf x}^{2}}{4\sigma_{0}^{2}}+ i{\bf k}.{\bf x}\right)
\label{initpack}
\end{equation}
where $\sigma_{0}$ is the initial width of the wave packet. The wave packet moves along the +ve $y-axis$ with the initial group velocity $v_{y}$
and the wave number $k_{y}=\frac{m v_{y}}{\hbar}$. Note that the initial spin state is given by  $\chi(t=0)=\frac{1}{\sqrt{2}}\left(\left|\uparrow\rangle_{z}+\right|\downarrow\rangle_{z}\right)$ where $\left|\uparrow\right\rangle_{z} $, $\left|\downarrow\right\rangle_{z}$ are eigenstates of the spin observable $\sigma_{z}$. The  inhomogeneous magnetic field (localised between $y=0$ and $y=d$) is directed along the $+ve$ $z-axis$. As the wave packet propagates through the SG magnet, in addition to the $+\widehat y -axis$ motion, the particles gain  velocity with the magnitude $v_{z}$ along $\widehat z-axis$  due to the interaction of their spins with the inhomogeneous magnetic field during the time interval $\tau$. 
 
Here the interaction Hamiltonian is $H_{int}=\mu\widehat \sigma.\textbf{B}$ where $\mu$ is the magnetic moment of the neutron, $\textbf{B}$ is the inhomogeneous 
magnetic field and $\widehat\sigma$ is the Pauli spin matrices vector. Then the time evolved total wave function  at $t=\tau$ ($\tau$ is taken to be the transit time of the peak of the wave packet within the SG magnetic field region) after the interaction of 
spins with the SG magnetic field is given by
\begin{eqnarray}
\nonumber
\Psi\left(\textbf{x},t=\tau\right) &=& \exp({-\frac{iH\tau}{\hbar}})\Psi(\textbf{x},t=0)\\
&=&\frac{1}{\sqrt{2}}\left[\psi_{+}(\textbf{x},\tau)\otimes\left|\uparrow\right\rangle_{z}+\psi_{-}(\textbf{x}, \tau)\otimes\left|\downarrow\right\rangle_{z}\right]
\label{timeevolved}
\end{eqnarray}
where $\psi_{+}\left({\bf x},\tau\right)$ and $\psi_{-}\left({\bf x},\tau\right)$ are the two components of the spinor $\psi=\left(\begin{array}{c}\begin{array}{c} \psi_{+}\\ \psi_{-}\end{array}\end{array}\right)$ which satisfies the Pauli equation. The inhomogeneous magnetic field is represented by ${\bf B}=(-bx,0,B_{0}+bz)$  satisfying the Maxwell equation ${ \nabla}.{\bf B}=0$. The two-component Pauli equation can then be written as two  coupled equations for $\psi_+$ and $\psi_-$, given by 
\begin{eqnarray}
i\hbar\frac{\partial {\psi}_+ }{\partial t} = -\frac{\hbar^{2}}{2m}\nabla^{2}{\psi}_{+}+\mu(B_0 + b z){\psi}_{+} -\mu b x {\psi}_- 
\nonumber\\
\label{coupledspinor}
\\
i\hbar\frac{\partial {\psi}_- }{\partial t} = -\frac{\hbar^{2}}{2m}\nabla^{2}{\psi}_{-}+\mu b x {\psi}_+ -\mu(B_0 + b z){\psi}_-
\nonumber
\end{eqnarray}
The coupling between the above two equations can be removed\cite{alstrom, cruz} using the condition $B_0\gg b\sigma_0$, whence one obtains the following decoupled equations given by
\begin{eqnarray}
i\hbar\frac{\partial\psi_{+}}{\partial t}&=&-\frac{\hbar^{2}}{2m}\nabla^{2}{\psi}_{+}+\mu\left(B_{0}+bz\right)\psi_{+} \nonumber\\
\label{decoupled}\\
i\hbar\frac{\partial\psi_{-}}{\partial t}&=&-\frac{\hbar^{2}}{2m}\nabla^{2}{\psi}_{-}-\mu\left(B_{0}+bz\right)\psi_{-}
\nonumber
\end{eqnarray}
The solutions of the above two equations are as follows
\begin{eqnarray}
\psi_{+}\left(\textbf{x};\tau\right)=\frac{1}{\left(2\pi s_{\tau}^{2}\right)^{\frac{3}{4}}}\exp\left[-\left\{ \frac{x^{2}+(y-v_{y}\tau)^{2}+(z-\frac{v_{z}\tau}{2})^{2}}{4\sigma_{0}s_{\tau}}\right\}\right]\nonumber\\
\times \exp\left[i\left\{-\Delta_{+}+ \left(y-\frac{v_{y}\tau}{2}\right)k_{y}+ k_{z}z\right\}\right]\nonumber\\
\label{solutions}
\\
\psi_{-}\left(\textbf{x};\tau\right)=\frac{1}{\left(2\pi s_{\tau}^{2}\right)^{\frac{3}{4}}}\exp\left[-\left\{ \frac{x^{2}+(y-v_{y}\tau)^{2}+(z+\frac{v_{z}\tau}{2})^{2}}{4\sigma_{0}s_{\tau}}\right\}\right]\nonumber\\
\times \exp \left[i\left\{-\Delta_{-}+ \left(y-\frac{v_{y}\tau}{2}\right)k_{y}-k_{z}z\right\} \right]
\nonumber
\end{eqnarray}
where $\Delta_{\pm}=\pm\frac{\mu B_{0}\tau}{\hbar}+\frac{m^{2} v_{z}^{2}\tau^{2}}{6\hbar^{2}}$,~~$v_{z}=\frac{\mu b\tau}{m}$,~~ $k_{z}=\frac{m v_{z}}{\hbar}$ and ~~$s_{t}=\sigma_{0}\left(1+\frac{i\hbar t}{2m\sigma_{0}^{2}}\right)$.
Here $\psi_{+}\left(\textbf{x},\tau\right)$ and $\psi_{-}\left(\textbf{x},
\tau\right)$ representing the spatial wave functions at $t=\tau$
correspond to the spin states $\left|\uparrow\right\rangle_{z} $ and $\left|\downarrow\right\rangle_{z} $ respectively, with the average momenta 
$\langle\widehat p\rangle_{\uparrow}$ and $\langle\widehat p\rangle_{\downarrow}$,
where $\langle\widehat p \rangle_{\uparrow\downarrow}=(0,mv_{y},\pm\mu b\tau)$.

Hence after emerging from the SG magnet, the particles corresponding to the wave function components
$\psi_{+}\left(\textbf{x},\tau\right)$ and $\psi_{-}\left(\textbf{x},\tau\right)$ move {\it freely}  along the respective directions  
$\widehat n_{+}=v_{y}\widehat j + \frac{\mu b\tau}{m}\widehat k$ and 
$\widehat n_{-}=v_{y}\widehat j - \frac{\mu b\tau}{m}\widehat k$ with 
the \emph{same} group velocity $v=\sqrt{v^{2}_{y}+(\frac{\mu b\tau}{m})^{2}}$ 
which is fixed by the relevant parameters of the SG setup and the initial velocity $(v_{y})$ of the peak of the 
wave packet. 

Now, the modulus of the inner product $I$ between the $\psi_{+}({\bf x},\tau)$ and 
$\psi_{-}({\bf x},\tau)$  is given by 
\begin{equation}
I=exp\left\{-\frac{\mu^{2}b^{2}\tau^{4}}{8 m^{2}\sigma_{0}^{2}}-\frac{2\mu^{2}b^{2}\tau^{2}\sigma_{0}^{2}}{\hbar^{2}}\right\}
\label{innerprod}
\end{equation}
that is necessarily \emph{zero} for the  ideal situation. 
This inner product is preserved for further time evolution during which
the freely evolving wave functions at a time $t_{1}$ \emph{after} emerging 
from SG setup are given by
\begin{eqnarray}
\fl
\psi_{+}({\bf x},t=\tau+t_{1})=\frac{1}{(2 \pi s^{2}_{t_{1}+\tau})^{3/4}}\exp\left[-\left\{\frac{x^{2}+\left(y-v_{y}(\tau+t_{1})\right)^{2}+\left(z-\frac{v_{z}\tau}{2}-v_{z}t_{1}\right)^2}{4 \sigma_0 s_{t_{1}+\tau}} \right\}\right]\nonumber\\
\times \exp\left[i\left\{-\Delta_{+} +k_{y}\left(y-\frac{v_{y}(\tau+t_{1})}{2}\right)+k_{z}(z-\frac{v_{z}t_{1}}{2})\right\}\right]\\
\fl \psi_{-}({\bf x},t=\tau+t_{1})=\frac{1}{(2\pi s^{2}_{t_{1}+\tau})^{3/4}} \exp\left[-\left\{\frac{x^{2}+(y-v_{y}(\tau+t_{1}))^{2}+(z+\frac{v_{z}\tau}{2}+v_{z}t_{1})^2}{4 \sigma_0 s_{t_{1}+\tau}} \right\}\right]\nonumber\\
\times \exp\left[i\left\{-\Delta_{-} +k_{y}\left(y-\frac{v_{z}(\tau+t_{1})}{2}\right)-k_{z}\left(z+\frac{v_{z}t_{1}}{2}\right)\right\}\right]\nonumber
\label{freewavefn}
\end{eqnarray}
where $s_{t_{1}+\tau}=\sigma_{0}\left(1+\frac{i\hbar (t_{1}+\tau)}{2m\sigma_{0}^{2}}\right)$.

Note that the wave packets $|\psi_{+}({\bf x},t=\tau)|^{2}$ and $|\psi_{-}({\bf x},t=\tau)|^{2}$ emerging from the SG magnet will move away from each other so that the position space overlap between these two wave packts will be changing with time. The position space overlap parameter $M(t)$ as defined in Eq.(7) is given by
\begin{equation}
M(t)= exp\left[-\frac{v_{z}^{2}(\tau+2t_{1})^{2}}{8\sigma_{\tau+t_{1}}^{2}}\right]
\end{equation}
where $\sigma_{\tau+t_{1}}$ is the width of the wave packet at the instant $\tau + t_{1}$. From the above equation, it follows that $M(t)$ \emph{saturates} to a \emph{time independent value} after a sufficiently large time, the saturated value being given by 
\begin{equation}
M_{s}=exp\left(-\frac{2 v_{z}^{2}m_{0}^{2} \sigma_{0}^{2}}{\hbar^{2}}\right)
\end{equation} 
\section*{Acknowledgements}
AKP acknowledges helpful discussions related to this work during his visits to the Perimeter Institute, Canada; Centre for Quantum Technologies, National University of Singapore; and Benasque Centre for Science, Spain. DH is grateful to Paul Davies and John Corbett for interactions that stimulated this work. DH thanks the Centre for Science and Consciousness, Kolkata for support. AKP acknowledges the Research Associateship of Bose Institute, Kolkata.

\end{document}